\documentclass[conference]{IEEEtran}
\usepackage{cite}
\usepackage{amsmath,amssymb}
\usepackage[]{algorithm2e}
\usepackage[standard]{ntheorem}
\usepackage{graphicx}
\usepackage[lofdepth,lotdepth]{subfig}

\newtheorem{claim}{Claim}

\begin{document}
\title{Online shortest paths with confidence intervals for routing in a time varying random network}

\author{\IEEEauthorblockN{St\'ephane Chr\'etien}
\IEEEauthorblockA{London Physical Laboratory, UK\\
Email: stephane.chretien@npl.co.uk}
\and
\IEEEauthorblockN{Christophe Guyeux}
\IEEEauthorblockA{Institut Femto-ST, UMR 6174 CNRS\\Universit\'e de Bourgogne Franche-Comt\'e, France\\
Email: christophe.guyeux@femto.fr\\}
%\and
%\IEEEauthorblockN{Osama Khalifa}
%\IEEEauthorblockA{Route Monkey\\
%\\
%\\
%}
}

\maketitle

\begin{abstract}
The increase in the world's population and rising standards of living is leading to an ever-increasing number of vehicles on the roads, and with it ever-increasing difficulties in traffic management. This traffic management in transport networks can be clearly optimized by using information and communication technologies referred as Intelligent Transport Systems (ITS). This management problem is usually reformulated as finding the shortest path in a time varying random graph. In this article, an online shortest path computation using stochastic gradient descent is proposed. This routing algorithm for ITS traffic management is based on the online Frank-Wolfe approach. Our improvement enables to find a confidence interval for the shortest path, by using the stochastic gradient algorithm for approximate Bayesian inference. The theory required to understand our approach is provided, as well as the implementation details. 
\end{abstract}

\IEEEpeerreviewmaketitle

\section{Introduction}

\subsection{Motivation}
Information and communication technologies have a great potential impact in the field of traffic and mobility management with a goal to achieve a safer and better optimized use of the transport network~\cite{elizondo}. Optimal routing is one of the key tasks involved in intelligent planning and an extensive body of work has been devoted to this topic since the 60's. The optimal routing problem is well known in the combinatorial optimisation community under the name of the \textit{shortest path} problem. Since the advent of Linear Programming in the 50's, a great number of optimisation problems related to logistics and planning have be shown to be special instances of convex optimisation. The \textit{shortest path problem}, in particular, is one of these problems for which a solution can be found in polynomial time using linear programming, and even faster algorithms are available such as the celebrated Dijkstra algorithm (or its generalisation, namely the A* algorithm).

In real world situations, however, there is never such a thing as a static network and the data available on the edge properties are evolving drastically in time. The cause of these variations are most often multiple: in e.g. transportation networks, a periodic behavior is definitely present in the traversal time of every route segment of the graph, because of the signal repeats from one day to the other. There might also be other periodicities such as depending on the day in the week, on the season, etc. There must also be various types of trends: linear or piecewise linear trends, jumps, exponential growth, etc. Jumps may be due to various kinds of disruptions. Piecewise linear signals can also occur in the presence of certain saturation phenomena. Exponential growth may be due to the increase of the population. On the top of this list of possible variations of the network properties come the fully random fluctuations, due to real time behavior on the roads. 

\subsection{Previous work on the stochastic shortest path problem}
In the case where the network is corrupted by random perturbations only, a very important work appeared in \cite{bertsekas1991analysis} where the problem is discussed in great mathematical details using a Markovian decision process approach. This model was also explored in \cite{psaraftis1993dynamic}. The dynamic programming approach was later refined in the subsequent paper \cite{polychronopoulos1993stochastic} where the computational complexity is significantly improved. The paper \cite{waller2002online} described a model with possible spatial and temporal correlations between the edges of the network. Starting from the results in 
\cite{waller2002online}, the interesting work \cite{fan2005shortest} described a efficient algorithm for solving the stochastic shortest path problem with correlations. A non-stationary version of this kind of model was proposed in \cite{gao2006optimal} where the problem is addressed using an adaptation of dynamic programming ideas.

Our approach to the shortest path problem was inspired by the recent breakthrough of \cite{abernethy2009competing} where the problem was recast as an online convex optimisation problem and the rate $m^{\frac32}\sqrt{T}$ was proved for a practically easy to implement algorithm. The online approach to the stochastic shortest path problem had long been interesting to the machine learning commmunity. The pioneering work by \cite{takimoto2002path} and 
then \cite{kalai2005efficient} set on the trend and proposed interesting and efficient algorithm. The partial information setting was studied in \cite{awerbuch2004adaptive} and superseeded by 
\cite{abernethy2009competing} using online convex optimisation techniques based on the linear programming representation of the shortest path problem based on the handy convex hull representation of $\mathcal P_{u,v}$. 

\subsection{Our contribution}
The previous works on the stochastic shortest path culminated in the discovery of the adaptive approach of \cite{abernethy2009competing} for the bandit setting with a controlled regret. A recent and elementary presentation of the online Frank-Wolfe approach to the problem is proposed in \cite[Section 7]{hazan2015introduction}. Our interest in the online approach was sparked by the difficulty to model the intricate compound of random and deterministic behavior which enter into the type of phenomena observable on real networks. 

The main ingredient we contribute to the online approach is a way to produce a confidence interval together with a shortest path. Our procedure is motivated by recent discoveries \cite{mandt2017stochastic} which prove that an appropriate rescaling of the Iterate Averaging Stochastic Gradient produces a sequence which is nearly Gaussian and has a distribution governed by the Bayesian posterior associated with the estimation problem, i.e. the shortest path in our setting. 

The main algorithm proposed in this paper combines the results of \cite{hazan2015introduction} with those of \cite{mandt2017stochastic} in order to propose an efficient way to simulate the distribution at time $T$ of the shortest path in a stochastic and potentially non-stationary (but smoothly evolving) environment. 

We demonstrate through numerical experiments the efficiency of our approach. 

\subsection{Plan of the paper}
The remainder of this article is organized as follows. The proposed algorithm is described in Section~\ref{sec:online} together with its practical implementation. Numerical simulations are then proposed in Section~\ref{sec:simuls}, for the sake of illustration. This research work ends by a conclusion section, in which the contribution is summarized and intended future work is outlined.

\section{Online routing}
\label{sec:online}
In this section, we describe our algorithm and its practical implementation.

\subsection{Online shortest path computation using stochastic gradient descent}

The shortest path problem in a time varying random graph is more complex to address than the standard shortest path problem in a deterministic graph. 

\subsubsection{Formulation}
We are given a graph $\mathcal G = (V,E)$ and a source-sink pair $(u,v) \in V^2$. Let $\mathcal P_{u,v}$ denote the set of all paths from $u$ to $v$. At each iteration $t$, the optimisation algorithm observes a set of weights $w_t \in \mathbb R_+^{\vert E\vert}$ and selects a path $p_t \in \mathcal P_{u,v}$. When the next iteration $t+1$ starts, a new set of observed weights $w_{t+1} \in \mathbb R_+^{\vert E\vert}$ is available and the decision maker observes a loss 
\begin{align}
    l_t & = \sum_{e \in p_t} \ w_t(e) 
\end{align}
associated with $p_t$. Our goal is to design a stochastic algorithm in order to perform iterative routing optimisation based on online sequentially updated information $w_t$ about the network state. 

One possible way to address this stochastic optimisation problem is to consider each path in $\mathcal P_{u,v}$ as an \textit{expert} and compare the loss of all experts in the pool in order to select the best among them. Unfortunately, the cardinality of $\mathcal P_{u,v}$ might be exponential and the resulting optimisation problem might end up being computationally intractable. 

Another, much more efficient option is to take the online convex optimisation approach of \cite{hazan2015introduction}. 

\subsubsection{The online Frank-Wolfe algorithm}
As is well known in combinatorial optimisation, the standard shortest path problem with edge weights $w_e$, $e\in E$, can be recast as a linear programming problem consisting in solving 
\begin{align}
    \min_{x \in \mathbb E^{\vert E\vert}} \ w^* x
\end{align}
subject to\footnote{here we make a slight abuse of notation by interpreting $\mathcal P_{u,v}$ as a set of binary vectors which are indicators of a path from $u$ to $v$, i.e. setting every component of $x$ to 1 when index by an edge on the path} $x \in \mathcal P_{u,v}$. 
The main idea is based on the fact that the $\mathcal P_{u,v}$ has the following convex hull: 
\begin{align}
    \sum_{c \in E} x_{u,c} & = 1 \label{spp1}\\
    \sum_{c \in E} x_{c,v} & = 1 \label{spp2}\\
    \forall c \in V\setminus \{u,v\}, \ \sum_{c': (c',c) \in E} x_{c',c} & = \sum_{c': (c,c')\in E} \ x_{c,c'} \label{spp3}\\
    \forall e \in E,\ x_e & \in [0,1].\label{spp4}
\end{align}

In our time varying random setting, we receive at each time $t$ a new value $w_t \in \mathbb R^{\vert E\vert}$ of the vector of weights and our goal is to achieve the minimal regret over a given time horizon $\{1,\ldots,T\}$, for a given $T \in \mathbb N_*$. The Online Conditional Gradient, aka Online Frank-Wolfe algorithm is a very efficient approach for doing this. The method is described in Algorithm \ref{OFW} below.

\begin{algorithm}%[H]
\SetAlgoLined
\KwResult{The path encoded in $x_T$}
 $x_1$ is a shortest path from $u$ to $v$, computed e.g. from averaged historical data\;
 \For{$t=1$ to $T$}{
   $y_{t+1} \leftarrow$ shortest path with edge cost vector 
   \begin{align}
       \eta \sum_{\tau =1}^t w_{\tau}+2(x_t-x_1)
   \end{align}
   set $x_{t+1} = \gamma_t \ x_t+(1-\gamma_t) \ y_{t+1}$ \;
 }
 \caption{The Online Frank-Wolfe Algorithm \cite[Section 7.5]{hazan2015introduction}\label{OFW}}
\end{algorithm}
The following result characterises the performance and adaptivity of this online method.

\begin{theorem}\cite[Theorem 7.2]{hazan2015introduction}
Let $D$ be the diameter of conv($\mathcal P_{u,v}$) and assume that $\Vert w_t\Vert_2 \le G$ for all $t\in \{1,\ldots,T\}$. The Online Frank-Wolfe Algorithm \ref{OFW} with parameters $\eta = G/(DT^{3/4})$
and $\gamma_t = \min \{1,2/t^{1/2}\}$ satisfies 
\begin{align}
    \sum_{t=1}^* w_t^*x_t-\min_{x \in \textrm{conv}(\mathcal P_{u,v})} \sum_{t=1}^T w_t^*x & \le 8DG \ T^{3/4}.
    \label{regr}
\end{align}

\end{theorem}
In other words, after dividing \eqref{regr} by $T$, this theorem says that the average loss suffered by the sequence of shortest paths is of the same order as the best average loss, i.e. the one suffered by one optimal path.

\subsection{Joint posterior distribution and confidence intervals}
The next question to be addressed is the one of providing a confidence interval for the shortest path. One idea is to apply the approach proposed in \cite{mandt2017stochastic} which consists in using the stochastic gradient algorithm for approximate Bayesian Inference. 

In the case where, instead of the Online Frank-Wolfe algorithm, one uses the Iterate Averaging Stochastic Gradient method defined in Algorithm \ref{iasg}.
\begin{algorithm}%[H]
\SetAlgoLined
\KwResult{The last iterate $x_T$}
 $x_1$ is an arbitrary initial point\;
 \For{$t=1$ to $T$}{
   $y_{t+1}  = y_t-\eta \nabla f_t(x_t)$ \\
   set $x_{t+1} = \gamma_t \ x_t+(1-\gamma_t) \ y_{t+1}$ \;
 }
 \caption{The Iterate Averaging Stochastic Gradient Algorithm \cite[Section6.1]{mandt2017stochastic}\label{iasg}}
\end{algorithm}
For this algorithm to be relevant, however, we need the function $f$ to be strongly convex.  Moreover, the vector  $\nabla f_t(x_t)$ will denote a centered stochastic perturbation of $\nabla f(x_t)$. 

For the Iterate Averaging Stochastic Gradient Algorithm, one obtains the following guarantees. 

\begin{claim}\cite[Based on Appendix G]{mandt2017stochastic}
Assume that 
\begin{align}
    \mathcal L_T (x) & = \frac1{T} \sum_{t=1}^T f_t(x)
\end{align}
is strongly convex. Take 
\begin{align}
    \gamma_t & = \frac{t}{t+1}.
    \label{gam}
\end{align} 
Then, we have 
\begin{align}
        \mathbb E\left[ x_{T}x_{T}^*\right] & \approx \frac1{\eta T} \left( 
    \Sigma(A^{-1})^*+A^{-1} \Sigma\right)
\end{align}
where $A$ is the Hessian of $\mathcal L_T$ at its minimiser.
\end{claim}
The accuracy of $\approx$ in the latter claim depends on how well $\mathcal L_T$ is approximated by $\frac12 x^tAx$ on the domain where the trajectory $(x_t)_{t=1,\ldots,T}$ will be wandering.

\subsection{Our new approach}
In our Online Frank-Wolfe setting, the problem is radically different from the Iterate Averaging Stochastic Gradient method, since the average cost $\mathcal L_T = \frac1{T} \sum_{t=1}^T \langle w_t,\cdot \rangle$ is linear, and therefore not strongly convex. However, we can use the same type of result for the output $x_T$ generated by the Online Frank-Wolfe Algorithm with 
$\gamma_t=t/(t+1)$ as in \eqref{gam}, and $\eta = 1/T$. Indeed, a similar diffusion-type approximation can be performed for the Online Frank Wolfe Algorithm as the one given in \cite{mandt2017stochastic} for the Iterate Averaging Stochastic Gradient if we replace the Ornstein-Uhlenbeck process by one similar to \cite{bubeck2015sampling}. The theoretical details will be given in an extended version of the present work.

\begin{algorithm}%[H]
\SetAlgoLined
\KwResult{The last iterate $x_T$}
\For{$l=1$ to $L$}{
Run Algorithm \ref{OFW} with $\gamma_t=t/(t+1)$ and $\eta = 1/T$ and return $x_T^{(l)}$\\
}
Compute the variance 
\begin{align}
    \hat \sigma & = \frac1L \sum_{l=1}^L \ \left(w^{(l)*}_Tx_{T}^{(l)}\right)^2- \left( \frac1{L}\sum_{l=1}^L w_T^{(l)*} x_{T}^{(l)}\right)^2.
    \end{align} \\
    Output the confidence interval 
    \begin{align}
        & \hspace{1cm} I_{\alpha} = \nonumber \\
        & 
        \left[\frac1{L}\sum_{l=1}^L w_T^{(l)*} x_{T}^{(l)}- u_{1-\alpha/2};\frac1{L}\sum_{l=1}^L w_T^{(l)*} x_{T}^{(l)}+u_{1-\alpha/2}\right]
    \end{align}
 \caption{The new sampling scheme }
\end{algorithm}

\section{Numerical simulation}
\label{sec:simuls}
\subsection{Simulation protocol}
We have designed an ad hoc transport network using the Python language~\cite{Rossum:1995:PRM:869369}. In this simulator, the number of crossroads has been set at various values, and each of them is linked to 2 to 5 other crossroads, according to a random draw. This design leads to a undirected graph, where nodes represent the crossroads, and edges are weighted according to the time needed between two nodes. Such weights are randomly picked in the real interval $[0;1]$ and updated at each time iterate, leading to a time varying random network. Such a dynamic graph has been implemented using the Networkx library~\cite{hagberg-2008-exploring}.

Algorithm~\ref{OFW} has then been computed, with the following parameters. $\eta$ has been set to: 
$$\eta = \dfrac{G}{DT^{\frac{3}{2}}},$$ 
where $G$ is equal to the square root of the number of edges, while $D$ is the largest distance between two shortest paths. $T$, for its part, is the number of steps of the algorithm, which has been set at 100. Finally, we have considered that:
$$\gamma_t=\dfrac{t}{t+1}.$$

At initialization stage, $w_0$, which is the vector of travel times (from size the number of edges) at the beginning of the simulation, which should in practice be equal to the averaged historical data, has been picked randomly. $y_1$ is the associated shortest path calculated, like all the other shortest paths, using the Dijkstra algorithm: the $i$-th component of $y_1$ is 1 if the $i$-th edge is in this shortest path, else it is equal to 0. Finally, $x_1$ has been set at $y_1$.

\subsection{Obtained results}

\begin{figure*}[ht] 
\begin{center} 
\subfloat[t=1]{ \includegraphics[width=0.4\textwidth]{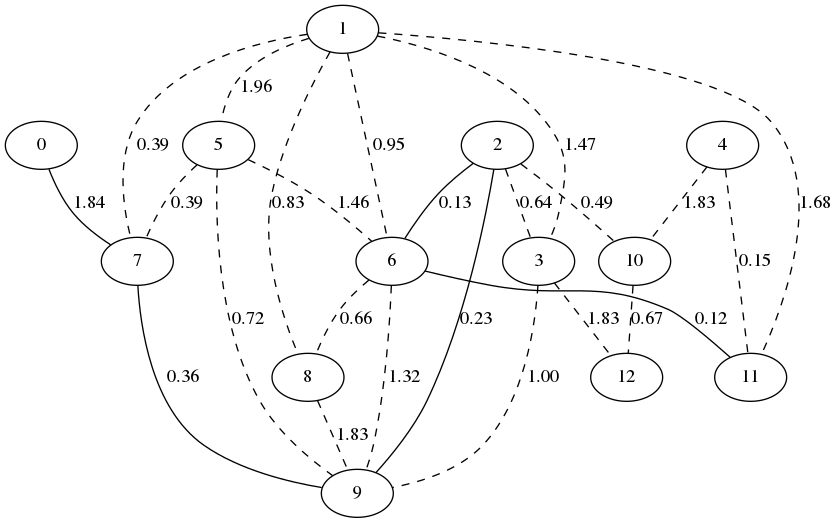}} 
\subfloat[t=5]{ \includegraphics[width=0.4\textwidth]{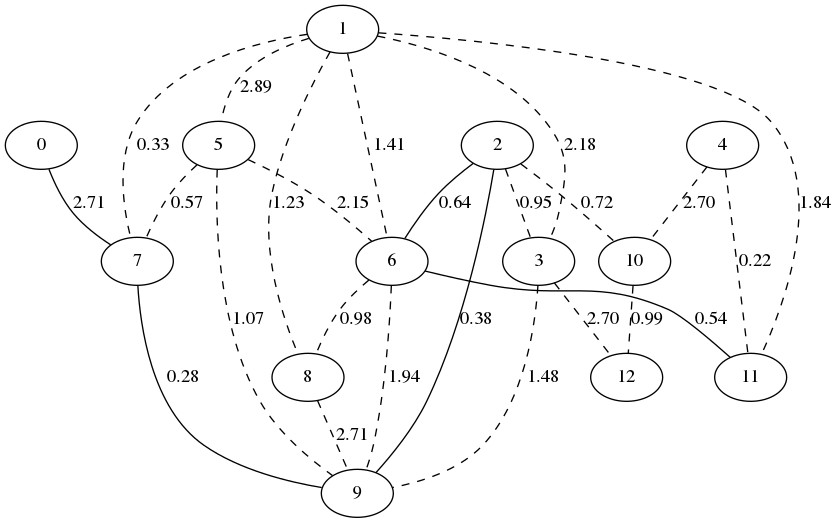}} \\

\subfloat[t=10]{ \includegraphics[width=0.4\textwidth]{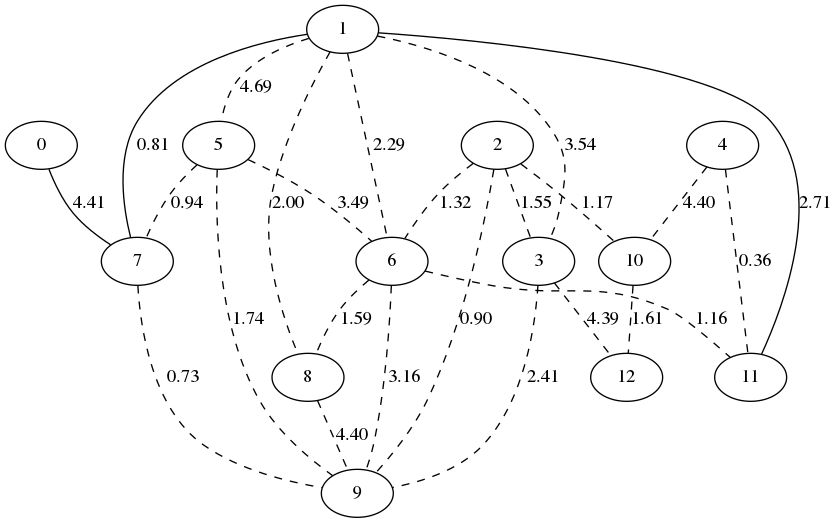}} 
\subfloat[t=15]{ \includegraphics[width=0.4\textwidth]{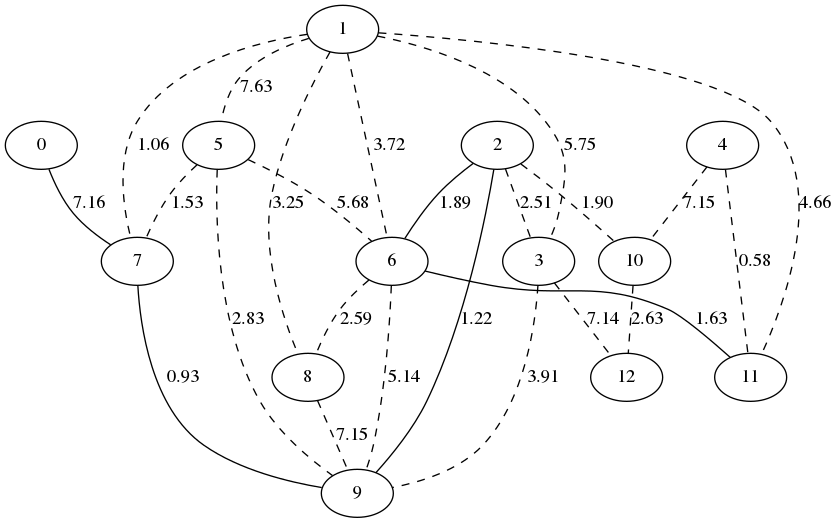}} \\

\subfloat[t=20]{ \includegraphics[width=0.4\textwidth]{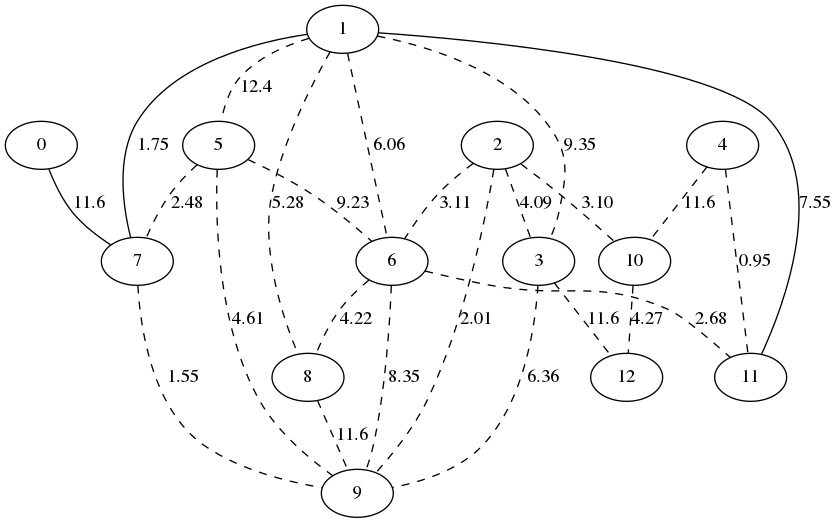}} 
\caption{The algorithm working on a small dynamic network (reaching 11 from 0)} \label{fig:renonculacees} 
\end{center} 
\end{figure*}

\begin{figure*}
    \centering
\subfloat[Cost evolution of shortest paths between two nodes over iterations, for different couples $(u,v)$ of source/sink vertices in the graph. Each color corresponds to a specific couple drawn uniformly at random.]{ \includegraphics[width=0.4\textwidth]{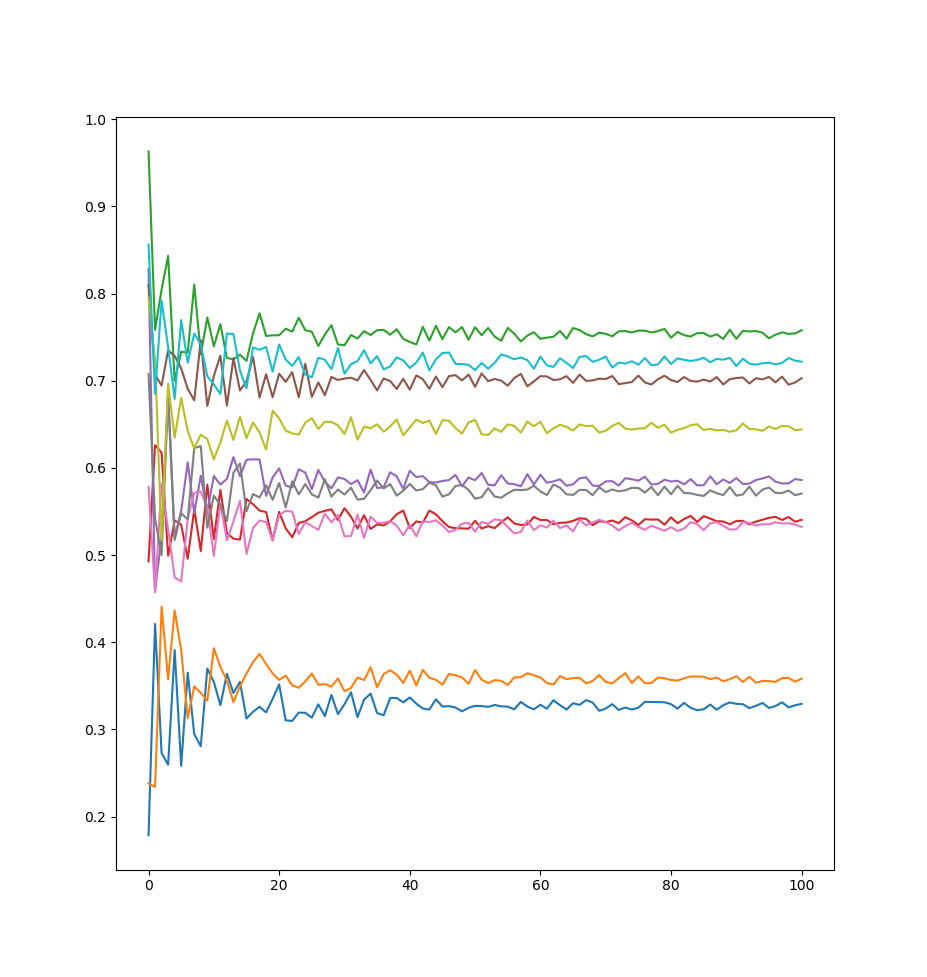}} \vspace{0.2cm}
\subfloat[Global computation time when the networks growth, mainly due to the Djikstra algorithm]{\includegraphics[width=0.5\textwidth]{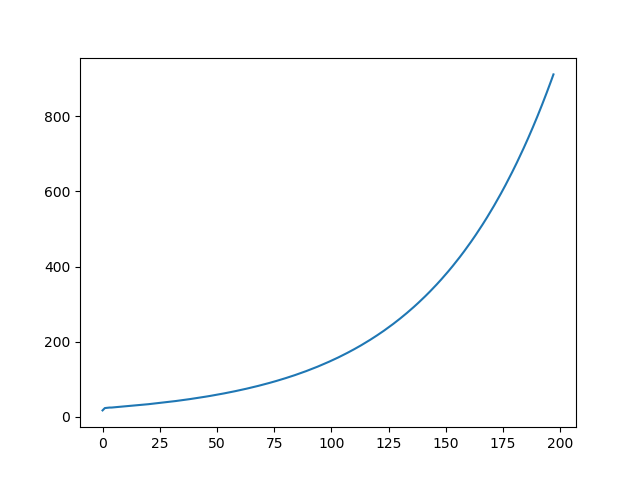}} 
    \caption{Convergence aspects of the proposal}
    \label{fig:iterations}
\end{figure*}

We have applied the proposed algorithm to various random networks, with a number of nodes respectively equal to $N=12, 100, 200,$ and $500$. The evolution of the shortest paths between any couple of nodes has been stored as movies, in which cumulative times between two locations are put on edges, while at each time only edges of the shortest path are not dotted. Examples of obtained results are depicted in Figure~\ref{fig:renonculacees} for 5 iteration steps and 12 nodes, while the mid term evolution of the algorithm to the shortest paths is presented in Figure~\ref{fig:iterations} for 10 couples of nodes within a network of 200 vertices.

\begin{figure*}[ht!]
    \centering
    \includegraphics[scale=0.3]{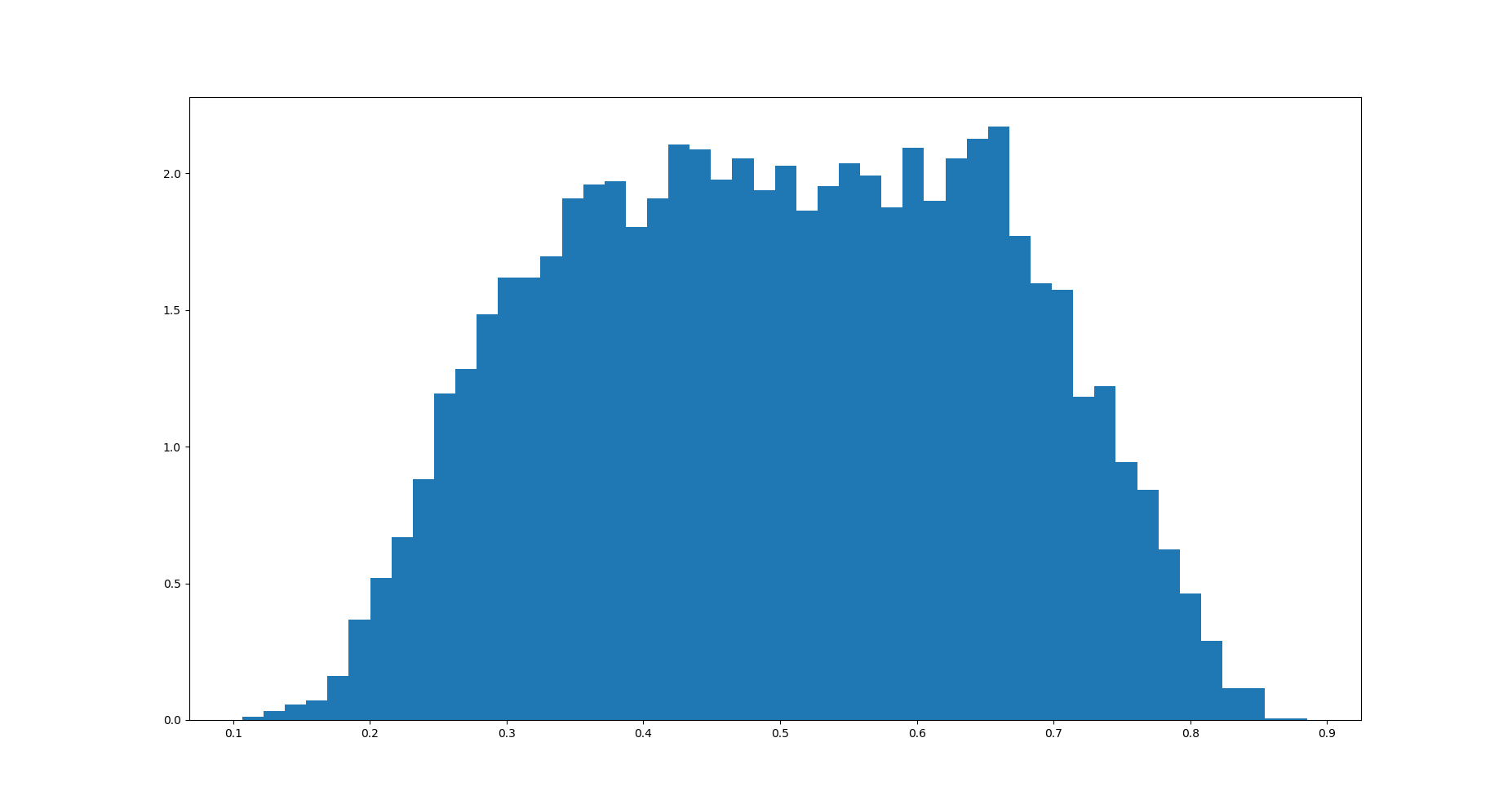}
    \caption{Distribution of obtained limits after convergence}
    \label{fig:global}
\end{figure*}

Finally, we have verified that the global computation cost of our approach is similar to the Djikstra algorithm (complexity of $O(n log(n))$, as can be seen in Figure~\ref{fig:global}. This result is encouraging, as we take place in a more difficult context of the shortest path problem, namely when the transport network evolves dynamically.

\section{Conclusion}
In this article, we proposed an online approach for the stochastic shortest path problem, which computes a confidence interval together with a shortest path. The approach is based on an appropriate rescalling of the Iterate Averaging Stochastic Gradient. Our main algorithm is a combination of the results of \cite{hazan2015introduction} and of \cite{mandt2017stochastic}, leading to an efficient way to simulate the distribution at time $T$ of the shortest path in a stochastic and potentially non-stationary (but smoothly evolving) environment. This algorithm has been implemented using the Networkx library of the Python language, and obtained results have been discussed.

In future work, the authors intention is to apply this online shortest path with confidence interval to situations different than traffic management in intelligent transport systems. One targeted field of research is the wireless sensor networks, especially the case of collaborative body sensor networks~\cite{bbdg17:ip,bbdgm16:ip} and of mobile ad-hoc networks (MANETs) in which the mobility of nodes implies an evolution in delay transmissions. Other directions of research encompass the use of new probability laws for edges, the introduction of correlations between two edges (or between two successive iterations in one given edge), and the variability in the number of nodes.

\section*{Acknowledgment}

C.G. acknowledges the financial support of the EU (Feder) and the Swiss Confederation within the framework of the Interreg France-Switzerland programme.
S.C. benefited from funding from the INNOVATE UK CAPITALS project http://gtr.rcuk.ac.uk/projects?ref=102618.

\bibliographystyle{plain}
\bibliography{bib}

\end{document}